\documentclass{IEEEtran}

\IEEEoverridecommandlockouts

\usepackage{listings}

\usepackage{algorithm}
\usepackage{algpseudocode}
\usepackage[
    colorlinks=true,
    urlcolor=blue,
    linkcolor=black,
    citecolor=black
]{hyperref}\usepackage{color}
\usepackage{amsmath}
\usepackage{amssymb}
\usepackage[thmmarks, amsmath]{ntheorem}
\usepackage{wrapfig}
\usepackage{epsfig}
\usepackage{epstopdf}
\usepackage{caption}
\usepackage{pdflscape}
\usepackage{makecell}
\usepackage[noadjust]{cite}
\usepackage{nicefrac}
\usepackage{stackengine}
\usepackage[cal=cm]{mathalfa}
\usepackage{float}
\usepackage{stfloats}
\usepackage{graphicx}
\newtheorem{definition}{Definition}

\usepackage[table,xcdraw]{xcolor}
\usepackage{acronym}
\usepackage{gensymb}
\usepackage{makecell}

\acrodef{ode}[ODE]{Ordinary Differential Equation}
\acrodef{dae}[DAE]{Differential Algebraic Equation}
\acrodef{rms}[RMS]{Root Mean Square}
\acrodef{emt}[EMT]{Electromagnetic Transient}
\acrodef{pf}[PF]{Participation Factor}
\acrodef{tf}[TF]{Transfer Function}
\acrodef{gnc}[GNC]{Generalised Nyquist Criterion}
\acrodef{vsc}[VSC]{Voltage Source Converter}
\acrodef{gfm}[GFM]{Grid-Forming}
\acrodef{gfl}[GFL]{Grid-Following}
\acrodef{sg}[SG]{Synchronous Generator}
\acrodef{stamp}[STAMP]{Small-Signal Toolbox for Analysis of Modern Power systems}
\acrodef{tds}[TDS]{Time Domain Simulation}
\acrodef{ibr}[IBR]{Inverter-Based Resources}
\acrodef{siso}[SISO]{Single-Input-Single-Output}
\acrodef{mimo}[MIMO]{Multiple-Input-Multiple-Output}
\acrodef{ipc}[IPC]{Interconnecting Power Converter}
\acrodef{ssa}[SSA]{Small-Signal Analysis}
\acrodef{hvdc}[HVDC]{High-Voltage-Direct-Current}
\acrodef{gnc}[GNC]{Generalized Nyquist Criterion}

\begin{document}
\title{A Matlab-based Toolbox for Automatic EMT Modeling and Small-Signal Stability Analysis of 
Modern Power Systems}
\author{\IEEEauthorblockN{Josep Ar\'{e}valo-Soler,~\IEEEmembership{Student Member,~IEEE}, Dionysios Moutevelis,~\IEEEmembership{Member,~IEEE}, Elia Mateu-Barriendos, \\ Onur Alican,~\IEEEmembership{Student Member,~IEEE}, Carlos Collados-Rodr\'{i}guez,  Marc Cheah-Mañe,~\IEEEmembership{Senior Member,~IEEE}, \\ Eduardo Prieto-Araujo,~\IEEEmembership{Senior Member,~IEEE}, and Oriol Gomis-Bellmunt,~\IEEEmembership{Fellow,~IEEE}}}
\vspace{-0.8cm}
\maketitle
\begin{abstract}
The intensive integration of power converters is changing the way that power systems operate, leading to the emergence of new types of dynamic phenomena and instabilities.
%
%
At the same time, converters act as an interface between traditional AC grids and their more recent DC counterparts, giving rise to hybrid AC/DC networks.
These conditions increase the necessity for stability analysis tools that can simultaneously account for the newly-introduced dynamic phenomena
and can also be applied for the stability study of hybrid networks.
%
%
This paper presents a Matlab-based toolbox for small-signal analysis of hybrid AC/DC power systems considering electromagnetic-transient (EMT) models.
The toolbox allows the automatized modeling of the system from the input data and offers options for modal, impedance and passivity analyses.
%
%
In the paper, the structure and internal processes of the toolbox are duly discussed, together with all its features, both main and complementary.
Its capabilities for stability analysis are demonstrated via comprehensive case studies of converter-based system of various size and topology.
%
%
%
%
%

\end{abstract}
\vspace{-0.2cm}
\section{Introduction}
\label{sec.introduction}
\subsection{Motivation}
Modern power systems are experiencing a deep transformation as a result of the massive integration of~\acp{ibr}.
%
Power electronics are being used massively in industrial and domestic loads, renewable generation units (mainly solar photovoltaics, onshore and offshore wind), Battery Energy Storage Systems (BESS), Flexible AC transmission systems (FACTS), and \ac{hvdc} transmission systems.
Power electronics-based systems have different characteristics compared to \acp{sg}, introducing dynamic phenomena on a much faster timescale~\cite{milano2018foundations}.
At the same time, they can act as an interface with DC networks, enabling the operation of hybrid AC/DC systems on both the transmission and the distribution level~\cite{van2016hvdc,harrison2024review}.
For these reasons, modern software for power system stability analysis needs to adapt to the evolving landscape introduced by~\acp{ibr}.
\subsection{Discussion on RMS vs EMT Modal Analysis}
In traditional power systems, the slow dynamics of \acp{sg} were considered the  most relevant, and potentially most hazardous for the proper power system operation~\cite{kundur2004definition}.
These dynamics were the driving force behind the most common undesirable phenomena, such as local and inter-area oscillations between generators and loss of synchronism during transients~\cite{kundur}.
%
The electromagnetic dynamics of the generators and the network, typically concerning frequencies much higher than the nominal one, were considered sufficiently damped and thus, not threatening to the stability of the system.
For this reason, these dynamics were simplified as algebraic equations in the models used for \ac{tds} and stability analysis, resulting in the classical \ac{dae} formulation for the power system.
This modeling approach is known as \ac{rms}, phasor-domain, or quasi-static phasor approach~\cite{lacerda2023phasor,lara2023revisiting}.

The integration of \acp{ibr} into the power system has triggered new instability phenomena at various timescales, referred to as converter-driven instabilities~\cite{hatziargyriou2020definition}.
%
%
These stability issues are commonly attributed to electromagnetic interactions between the power converter controllers, the converter filters, and the electrical grid components, such as lines, transformers, or passive filters.
For this reason, detailed modeling of \acp{ibr} is required~\cite{paolone2020fundamentals}.
In this context, the standard quasi-stationary approximation for network and electrical elements is no longer accurate, requiring a fully \ac{ode} representation for both network and shunt devices.
This approach is termed~\ac{emt} modeling and recently has been the standard approach for~\ac{tds} studies on the device level for \ac{ibr} integration studies~\cite{lara2023revisiting}.
It should be noted that in the literature, different levels of detail exist between \ac{rms} and \ac{emt} modeling~\cite{lacerda2023phasor,lara2023revisiting}.
%
%

\ac{ssa}
is the most frequently used method for the stability analysis of power systems during small disturbances around their nominal equilibrium point~\cite{kundur,milano2020eigenvalue}.
The majority of commercially available software tools for power system stability analysis use \ac{rms} models for \ac{ssa}, while the adoption of \ac{emt} models for stability oriented, system level~\ac{ssa} is still largely underdeveloped.
In the scope of this work, small-signal analysis performed with \ac{rms} models will be referred to as \ac{ssa}-\ac{rms}, while the one considering \ac{emt} models will be referred to as \ac{ssa}-\ac{emt}.

\subsection{Literature Review}
Software solutions for power system simulation and stability analysis are typically distinguished between commercial and open-source~\cite{milano2005open}.
Commercial software packages widely used in industry include PSCAD,  PSS/E and Digsilent Powerfactory~\cite{PSCAD2023,PSSE2023,DIgSILENT2023}.
Their main scope regards \ac{tds}, for which a wide variety of models, ranging from \ac{rms} and \ac{emt} to switching models, is included in their integrated libraries.
However, with regards to \ac{ssa}, the included models for the transmission network revert to an~\ac{rms} formulation, thus neglecting high frequency modes that may introduce adverse interactions for the power system safe operation.

Open-source software packages are usually developed for research and educational purposes.
While they lack the warranties and elaborate user interface of commercial options, their libraries and algorithmic routines are typically available for inspection and modification, allowing more flexibility to individual users.
Several research software projects regarding power system simulation and stability analysis have been  presented in recent years~\cite{aristidou2013dynamic,xiong2023paraemt,milano2005open,milano2013python,cui2020hybrid,serrano2022stability,kelada2025revisiting}.
%
Software packages RAMSES and ParaEMT offer solutions for efficient~\ac{tds} of large power systems using parallelization techniques, based on \ac{rms} and \ac{emt} modeling of the power system, respectively~\cite{aristidou2013dynamic,xiong2023paraemt}.
However, no option for~\ac{ssa} is included.
The PSAT, implemented in Matlab environment and its follow-up Dome, implemented in Python, offer features for~\ac{tds},~\ac{ssa}, optimal power flow and continuation power flow~\cite{milano2005open,milano2013python}.
ANDES is another Python-based library that can perform power flow calculations,~\ac{tds} and~\ac{ssa}~\cite{cui2020hybrid}.
%
%
However, despite the option found in Dome for ad hoc inclusion of selected network dynamics~\cite{milano2016semi}, all three of the these software packages are based on an~\ac{rms} formulation of the power system, preserving in principle the dynamic properties of the various shunt devices connected to the network, but neglecting the dynamic effect of the network itself.

Recently, software tools for the \ac{emt}-\ac{ssa} have been presented~\cite{serrano2022stability,kelada2025revisiting}.
CSTEP, developed in Matlab environment, models the electrical element of the network as dynamic variables and additionally includes a feature for automatically removing redundant variables, for example variables representing current of inductors connected in series~\cite{serrano2022stability}.
However, it relies on a symbolic variable internal engine which may limit the scalability of the \ac{ssa} for very large systems.
Reference~\cite{kelada2025revisiting} also introduced a Matlab-based, \ac{ssa} toolbox, considering important effects such as transmission lines and \ac{sg} stator dynamics.
However neither of the two \ac{emt}-based, \ac{ssa} tools offer the capability of including AC/DC hybrid systems.

\subsection{Contributions}
\begin{figure}[!t]
\vspace{-0.2cm}
\centering\includegraphics[width=0.49\textwidth]{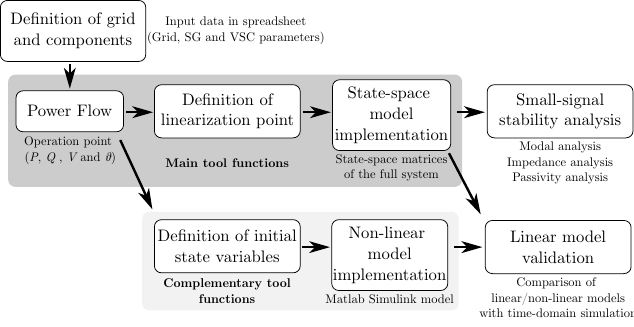}
\vspace{-0.2cm}
\caption{Flowchart with STAMP structure.}
\vspace{-0.3cm}
\label{fig.overview}
\end{figure} 
This paper introduces the~\ac{stamp}, a novel Matlab-based toolbox for automatic \ac{emt} modeling and \ac{ssa} of hybrid AC/DC power systems.
The tool is developed in Matlab, chosen for its matrix-oriented programming, robust plotting capabilities, and graphical environment provided by Simulink.
While the main scope of \ac{stamp} revolves around \ac{ssa} and not on \ac{tds} of large power systems, non-linear power systems models are also automatically generated as part of its processes.
The main value of these non-linear models is the validation of the linear models that are used for the \ac{ssa}.

The main contributions of \ac{stamp} are summarized as follows:
\begin{itemize}
    \item A comprehensive methodology for state-space modeling of hybrid AC/DC networks
    for \ac{ssa}-\ac{emt} purposes.
    \item An automatic process to generate linear models for \ac{ssa}-\ac{emt} in large-scale networks with multiple \acp{ibr}.
\end{itemize}
Other supplementary features of the toolbox include:
\begin{itemize}
    \item A modular architecture which allows the seamless incorporation of new, user-defined elements, either in analytic or numerical form.
    \item A comprehensive library of predefined linear models for various elements, including \acp{sg}, \acp{ibr} and \ac{hvdc} converters. Control options for both \ac{gfm} and \ac{gfl} operation modes are considered.
    \item Initialization routines for the non-linear models so that the~\acp{tds} start from their predefined equilibrium point.
    \item Clear network model visualization stemming from the block-based, graphic interface of Simulink.
    \item Various options for \ac{ssa}, including eigenvalue parametric sensitivity scanning, impedance and passivity analysis.
\end{itemize}

To the extent of the authors' knowledge, \ac{stamp} is the first publicly available software tool for \ac{ssa}-\ac{emt}
of large scale hybrid AC/DC power systems, aimed for research and educational purposes~\cite{STAMP_Github}.
Its general structure is depicted in Fig. \ref{fig.overview}.
In the following, the core functionalities of~\ac{stamp} are presented and its capability for \ac{ssa} of converter-based systems is demonstrated through case studies based on power systems of different size and topology.
\subsection{Paper Organization}
The organization of the rest of the paper is the following. Section~\ref{sec.modeling} presents the~\ac{ode}-based modeling approach of \ac{stamp}, while Section~\ref{sec.ssmethods} presents the theoretical background behind the \ac{ssa} options included in the tool.
Section~\ref{sec.implementation} explains the implementation aspects of the linear, state-space models for the \ac{ssa}-\ac{emt}, with emphasis on the AC and DC network modeling, which is the main novelty compared to \ac{rms} approaches.
Section~\ref{sec.complementary_features} details 
some of the outlying features of \ac{stamp}.
Finally, Section~\ref{sec.case_studies} presents the case studies showcasing the modeling and analysis capabilities of the tool, while Section~\ref{sec.conclusion} concludes the paper.
\section{Small-Signal Modeling}
\label{sec.modeling}

In this section, the modeling approach considered in the \ac{stamp} for \ac{ssa} are introduced.

\subsection{Power System Dynamic Modeling}
\label{sec.definition_general}

A power system can be modeled as a set of \acp{ode}~\cite{kundur}:
\begin{equation}  
\label{eq:ode_all}
\begin{aligned}
\frac{d \boldsymbol{x}}{dt}
=
\boldsymbol{f(x,u)},
\end{aligned}
\end{equation}
%
%
where $\boldsymbol{x}$ and $\boldsymbol{u}$ are the column vectors of state and input variables, respectively, and $\boldsymbol{f}$ is a non-linear function. These \acp{ode} can be split in two groups:
\begin{equation}  
\label{eq:ode_split}
\begin{aligned}
\frac{d \boldsymbol{x_d}}{dt}
=
\boldsymbol{f_d}( \boldsymbol{x},\boldsymbol{u}),\quad
\frac{d \boldsymbol{x_g}}{dt}
=
\boldsymbol{f_g}(\boldsymbol{x},\boldsymbol{u}),
\end{aligned}
\end{equation}
where $\boldsymbol{x}=[\boldsymbol{x_d} \; \boldsymbol{x_g}]^\intercal$ and $\boldsymbol{f}=[\boldsymbol{f_d} \; \boldsymbol{f_g}]^\intercal$ with subindex $d$ referring to shunt devices (generators, loads, etc.) and subindex $g$ referring to grid elements.
Although generally $\boldsymbol{x_g}$ contains variables related to the transmission grid, e.g., bus voltage magnitudes and angles as well as the line currents, it can also include other device-related, internal electrical variables with fast dynamics, e.g., \ac{sg} stator and converter filters dynamics.


\subsection{\ac{rms} power system modeling}
\label{sec.rms_modeling}
By setting $d\boldsymbol{x_g}/dt=0$ in \eqref{eq:ode_split}, the classical~\ac{dae} representation of power systems is obtained:
\begin{equation}  
\label{eq:dae_general}
\begin{aligned}
\frac{d \boldsymbol{x_d}}{dt}
=
\boldsymbol{f_d}( \boldsymbol{x},\boldsymbol{u}),\quad
0
=
\boldsymbol{f_g}(\boldsymbol{x},\boldsymbol{u}),
\end{aligned}
\end{equation}
with $\boldsymbol{f_g}$ in this formulation representing the algebraic constraints of the system and $\boldsymbol{x_g}$ being algebraic variables.
The \ac{dae} representation, also known as~\ac{rms} model, is traditionally used for the study of electromechanical and slow control dynamics up to tenths of Hz.
The linear model of the system can be obtain from~\eqref{eq:dae_general} following the process described in~\cite{moutevelis2024modal,milano2010power}.
%
%
\subsection{\ac{emt} linear modeling}
\label{sec.ssmodels}
By 
considering $d\boldsymbol{x_d}/dt\neq0$ and $d\boldsymbol{x_g}/dt\neq0$ in \eqref{eq:ode_split}, the full dynamics of the system, including the transmission grid dynamics, are included in the formulation.
The linear models for \ac{ssa}-\ac{emt} are obtained by linearizing the complete \acp{ode} from \eqref{eq:ode_all}.
In particular, by applying a first-order Taylor series expansion to~\eqref{eq:ode_all} and selecting the output variables and grouping them in the column vector $\boldsymbol{y}$, a linear state-space model of the system is obtained as follows~\cite{ogata_book}:
\begin{equation}
\label{eq:linear_ode}
\begin{aligned}
\frac{d \Delta \boldsymbol{x}}{dt}=\boldsymbol{A} \Delta \boldsymbol{x}+ \boldsymbol{B} \Delta \boldsymbol{u},
\\   \Delta \boldsymbol{y}=\boldsymbol{C} \Delta \boldsymbol{x} + \boldsymbol{D} \Delta \boldsymbol{u},
\end{aligned}     
\end{equation}
where, $\Delta$ stands for ``small perturbation'', $\boldsymbol{A}$, $\boldsymbol{B}$, $\boldsymbol{C}$, and $\boldsymbol{D}$ are the state, input, output and feed-forward matrices, respectively.
The formulation in~\eqref{eq:linear_ode} captures a wider range of dynamics compared to~\eqref{eq:dae_general}, and it is the modeling method implemented in \ac{stamp}.
\subsection{Reference Frame Transformations}
\label{sec.reference_frames}
%
%
%
%
%
Three-phase balanced power systems can be expressed in $dq$ or $qd$ rotating reference frames after applying a Park transformation, with the latter being the preferred option in \ac{stamp}~\cite{egea2012active}.
%
%
Representation in rotating reference frames allows linearizing around constant operation points, simplifying the \ac{ssa}.
Global and local $qd$ reference frames are defined for the interconnection of several devices to the electrical grid.
The global reference frame refers to a common $QD$ reference frame, defined by the selected slack device of the electrical grid, typically an ideal voltage source or a generator device.
%
%
The local reference frame is the $qd-i$ reference frame from each device $i$, usually \ac{sg} or \ac{vsc} devices.
%
%

In order to transform the vector of signals $\boldsymbol{e}_{qd-i}=[e_q \; e_d]^\intercal$ from the local reference frame $qd-i$ of each device $i$ to a vector $\boldsymbol{e}_{QD}=[e_Q \; e_D]^\intercal$ expressed in the global reference frame $QD$, rotation matrices are used as in~\cite{moutevelis2022bifurcation,pogaku2007modeling}:
%
%
\begin{equation}
\label{eq:transformation}
\boldsymbol{e}_{QD}
={
\begin{bmatrix}
\cos \delta_i & -\sin \delta_i\\
\sin\delta_i & \;\;\;\cos \delta_i
\end{bmatrix}}
\boldsymbol{e}_{qd-i},
\end{equation}
\noindent where $\delta_i$ is the angle between the reference frames and is defined by the synchronization dynamics of each device. 

%

\subsection{Impedance representation}
\label{sec.impmodel}

For \ac{siso} systems, the connection between the state-space representation and the \ac{tf} between the system input and output in the Laplace domain can be derived as~\cite{ogata_book}:
\begin{equation}
\label{eq.ABCD2tf}
     G(s)
     =
     \frac{\Delta y(s)}{\Delta u(s)}
     =
     \boldsymbol{C}(s\boldsymbol{I}-\boldsymbol{A})^{-1}\boldsymbol{B}+\boldsymbol{D},
\end{equation}
where $s$ is the Laplace operator and $^{-1}$ is the inverse matrix operator.
%
When the expression~\eqref{eq.ABCD2tf} considers voltages and currents as inputs and outputs then the \ac{tf} represents a impedance  $Z(s)$ or an admittance $Y(s)$. The impedance or admittance are evaluated for a single device or from a bus of a power system. 

Three-phase balanced power systems usually consider $qd$ components, which results in a \ac{mimo} system. Then, the scalar \acp{tf} $Z(s)$ and $Y(s)$ become $2\times2$ matrices $\boldsymbol{Z}(s)$ and $\boldsymbol{Y}(s)$, respectively:
\begin{align}
\boldsymbol{Z}(s)
=
\boldsymbol{Z_{qd}}(s)
=
\begin{bmatrix} 
Z_{qq}(s) & Z_{qd}(s) \\
Z_{dq}(s) & Z_{dd}(s) 
\end {bmatrix}, \\
\;
\boldsymbol{Y}(s)
= 
\boldsymbol{Y_{qd}}(s)
=
\begin{bmatrix} 
Y_{qq}(s) & Y_{qd}(s) \\
Y_{dq}(s) & Y_{dd}(s) 
\end {bmatrix}.
\end{align}
%
%
The impedance and admittance \ac{tf} matrices can also be expressed in the positive-negative sequence frame, which in \ac{stamp} is implemented via the transformation~\cite{rygg2016modified}:
\begin{align}
\label{eq.qd2pn}
\boldsymbol{Z_{pn}}(s)
=
\boldsymbol{T}
\boldsymbol{Z_{qd}}(s)
\boldsymbol{T^{-1}}
,
\;
\boldsymbol{T}
= 
\frac{1}{\sqrt{2}}
\begin{bmatrix} 
1 & -j \\
1 &  j
\end{bmatrix}.
\end{align}


\section{Small-signal analysis methods}
\label{sec.ssmethods}
Small-signal analysis for power systems can be mainly divided into modal analysis, resulting from state-space representations, and frequency-domain analysis, resulting mainly from impedance representations.
In particular, \ac{stamp} considers the following methods:
\begin{itemize}
    \item Modal analysis (eigenvalue- and participation factor-based).
    \item Impedance analysis.
    \item Passivity analysis.
\end{itemize}

The necessary metrics for these methods, which are detailed in the following subsections, can all be computed from the state-space representation resulting from the two linear modeling 
approaches presented in Section \ref{sec.modeling}.
In order to 
formally establish the linear analysis variations based on each modeling approach, the following two definitions are introduced.
%
These definitions serve to further highlight the differences resulting from the selected power system representation, not only with regards to non-linear \ac{tds}, but also for linear \ac{ssa}.

\begin{definition}
(\textbf{Small-Signal Analysis with Root Mean Square models (SSA-RMS)}) linearizes power system dynamics around an operating point while neglecting fast electromagnetic transients, i.e., $d\boldsymbol{x_g}/dt=0$, which results in a DAE formulation. This is equivalent to linearizing model~\eqref{eq:dae_general}. 
\end{definition}

\begin{definition}
(\textbf{Small-Signal Analysis with Electromagnetic Transient models (SSA-EMT)}) linearizes power system dynamics around an operating point while considering fast electromagnetic transients, i.e., $d\boldsymbol{x_g}/dt\neq0$, and  $d\boldsymbol{x_d}/dt\neq0$, which results in a fully ODE formulation. This equivalent to linearizing model~\eqref{eq:ode_all}.
\end{definition}
The \ac{stamp} implements automatized \ac{ssa} considering an \ac{ssa}-\ac{emt} approach representing dynamics up to a few kilohertz, but neglecting the converter switching dynamics.
Despite not lying within its main scope, \ac{ssa}-\ac{rms} models can also be derived from the \ac{ssa}-\ac{rms} by applying various reduction techniques, for example Kron reduction~\cite{dorfler2012kron}.
\subsection{Modal Analysis}
\label{sec.modal}
The analysis of matrix $\boldsymbol{A}$ from \eqref{eq:linear_ode} provides several relevant indices for the stability assessment with the most notable ones being the eigenvalues and the \acp{pf}~\cite{kundur}.
The eigenvalues $\lambda_i$ are calculated as the solutions of equation~\cite{milano2020eigenvalue}:
\begin{equation}
\label{eq.eigen}
     |\boldsymbol{A} - \lambda_i \boldsymbol{I_i}|=0,
\end{equation}
where $|\cdot|$ is the determinant of a matrix, $i$ is the order of matrix $\boldsymbol{A}$ and $\boldsymbol{I_i}$ is a unity matrix of order $i$.
The sign of the eigenvalues' real part determines the stability of the system, with the dynamic system being asymptotically stable if all eigenvalues have a negative real part. In case of oscillatory modes, which are represented as complex conjugate eigenvalues, the frequency and damping ratio terms can also be obtained.

The \acp{pf} $p_{kj}$
relate the different modes of the system~(represented via the eigenvalues~$\lambda_i$) to the states of the linearised system $\Delta \boldsymbol{ x}$~\cite{moutevelis2024modal}, and are calculated as~\cite{kundur}:
\begin{equation}
\label{eq.pfs}
    p_{kj}=\psi_{jk} \phi_{kj},
\end{equation}
where $\psi_{jk}$ and $\phi_{kj}$ are the elements of the left and right eigenvectors, respectively, of matrix $\boldsymbol{A}$.
In \ac{stamp}, the \acp{pf} are normalized based on the maximum \ac{pf} value for every state, namely: 
\begin{equation}
\label{eq.pfs_normalised}
    \hat{p}_{kj}=|p_{kj} \frac{1}{\max\limits_{k}(|p_{kj}|)}|,
\end{equation}
where $\hat{p}_{kj}$ are the normalised \acp{pf}.
\subsection{Impedance Analysis}
\label{sec.impedance}
The frequency-domain analysis from impedance representation is typically considered for power system analysis with converters. The frequency-domain characteristics of the impedance can be obtained with frequency scan techniques or directly as a \ac{tf} from the state-space model, as indicated in \eqref{eq.ABCD2tf}. The latter is the option considered in \ac{stamp}.

The frequency-domain analysis provides information regarding the dynamic interactions between devices and the grid, and can be used to assess the overall stability of interconnected systems. Several methods have been proposed in the literature for assessing stability using impedance representations, with the most widely adopted being the impedance-based stability analysis, also known as impedance ratio analysis~\cite{sun2011impedance}.

This method partitions the power system into two subsystems: the device under study, with admittance $\boldsymbol{Y_d}(s)$, and the remainder of the grid, with impedance $\boldsymbol{Z_g}(s)$. The resulting impedance ratio $\boldsymbol{Z_g}(s)\cdot\boldsymbol{Y_d}(s)$, also referred to as the minor loop gain, serves as the system's open-loop \ac{tf} and is used to assess stability of the system's closed-loop. 
First, potential resonances are identified at the frequencies where the impedance \ac{tf}s exhibits sharp peaks. Then, the stability can be evaluated by applying the Nyquist Criterion to the impedance ratio ~\cite{amin2019nyquist}.
For the cases of three-phase power systems formulated in rotating reference frames, a \ac{mimo} variation of the Nyquist Criterion, called the \ac{gnc}, is used~\cite{amin2019nyquist,moutevelis2024virtual}.
It must be noted that to apply this approach, both subsystems must be individually stable.

%
\subsection{Passivity Analysis}
\label{sec.passivity}
Assessing the passivity property of a dynamic system is another way to derive information regarding the dynamic operation of a system~\cite{harnefors2015passivity}.
Specifically, the interconnection of passive elements is guaranteed to be stable, since all the dynamic elements of the network exhibit damping capability of destabilizing oscillations.
%
The passivity of a dynamic device can be assessed across a desired frequency range $\omega_{rng}$ through its impedance \ac{tf} $\boldsymbol{Z}(s)$~(equivalently through its admittance  \ac{tf} $\boldsymbol{Y}(s)$).
This is achieved by checking the following conditions~\cite{arevalo2024converter,ortega2021pid}:
\begin{equation}
  \label{eq:passivity}
  \begin{aligned}
    & \bullet \;
    \boldsymbol{Z}(s) \; \text{is stable.}
    \\
    & \bullet \;
    \boldsymbol{H}(j\omega)
    =
    \boldsymbol{Z}(j\omega) + \boldsymbol{Z}^{H}(j\omega) \geq 0, \forall \omega \in \omega_{rng},  
  \end{aligned}
\end{equation}
where $^H$ is the Hermitian conjugate operator and the inequality represents a positive semi-definite matrix.
In \ac{stamp}, condition~\eqref{eq:passivity} is checked by calculating the minimum eigenvalue $\lambda_{min}$ of matrix $\boldsymbol{H}$ across the entirety of the frequency spectrum of interest.
For the regions that $\lambda_{min}$ is positive, the device is passive.


\section{Implementation of state-space model}
\label{sec.implementation}
\begin{figure}[!t]
    \centering    
    \includegraphics[width=1\columnwidth]{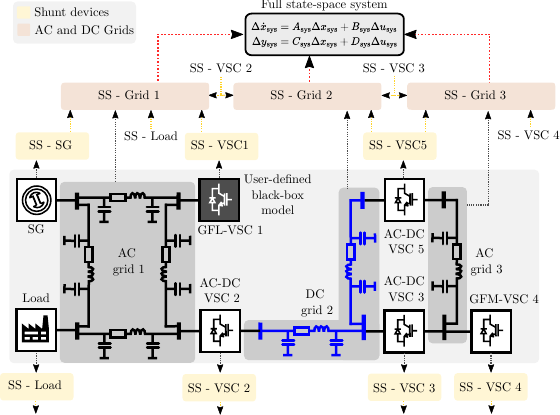}
    \caption{Example of state-space model interconnections.}
    \label{fig:ss_implementation}
\end{figure}
%
%
%
The process that \ac{stamp} uses for the derivation of the complete state-space, linear model relies on the division of the full system into smaller subsystems and their subsequent synthesis. Then, two main steps are followed:
\begin{enumerate}
    \item The state-space model of each individual component (e.g., grid element, device, or control block) is calculated as a separate subsystem.
    \item The different subsystems are arranged and interconnected based on their respective inputs and outputs, considering the process described in~\cite{nasa,gaba1988comparative}. This subsystem interconnection creates a meshed structure of state-space subsystems that represents the linear dynamic behavior of the full system, as illustrated in Fig.~\ref{fig:ss_implementation}.
\end{enumerate}

%
%
The interconnection process is implemented in \ac{stamp} using the Matlab command \textit{connect} from the Signal Processing Toolbox~\cite{MATLAB2023b_SPT}.
%
%
Each individual linear subsystem is grouped within higher level subsystems that serve as an intermediate level before the full model interconnection.
Specifically, these groups are the following:
%
%
\begin{itemize}
    \item AC grid, which includes AC lines and transformers. 
    \item DC grid, which includes DC lines.
    \item Individual shunt devices, such as generators, loads, and other grid shunt elements (e.g., reactive power compensation components) that are connected to the AC and DC grids.
\end{itemize}
The introduction of AC and DC grid subsystems is especially relevant for \ac{emt} stability studies and represents one of the main contributions of \ac{stamp}.
For this reason, in the following subsections, more details regarding the state-space model derivation of each group are provided.
%
\subsection{AC grid model}
\begin{figure}[!t]
    \centering    
    \includegraphics{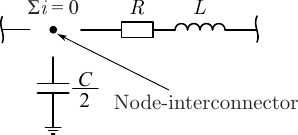}
    \caption{$\pi$-section model divided in different subsystems}
    \label{fig:pi_separat}
\end{figure}
%

%
%
In this subsection, the process to automatically incorporate the dynamic model of lines, modeled as $\pi$-section circuits, into the state-space model is explained.
The approach is not limited to this type of line model and can also be applied to other types of two-port systems, such as $RL$ lines and transformers.
However, in order to apply this method, the connection of at least one shunt capacitor from a $\pi$-section line modeled is assumed at each system bus.

The implementation of the AC grid linear model relies on the division of the $\pi$-section state-space model into three different subsystems, as depicted in Fig.~\ref{fig:pi_separat}.
%
These subsystems include the equivalent line capacitance $C$, the equivalent $RL$-line section, and a \textit{node-interconnector} element.
%
In particular, the node-interconnector element represents the application of the Kirchhoff Current Law for each bus and enables the integration of all the elements of the full system.
The associate model of the node-interconnector element is represented by algebraic equations contained in matrix $\boldsymbol{D}$ of its state-space representation.
%
%
This matrix can be updated to incorporate the current injections from the other AC grid elements.
A formal definition for the node-interconnector elements is provided as follows.
\begin{definition}
(\textbf{Node-interconnector}) It is an interconnector element formulated with a state space representation. This model consists of:

\begin{itemize}
    \item \textbf{State matrix}: Is empty as it does not include state variables.
    \item \textbf{Input vector}: Defined by all currents flowing into the \textit{node-interconnector}.
    \item \textbf{Output vector}: \( \Delta y(t) = [\Delta i^q_{c}, \Delta i^d_{c}] \), where \( i^{q,d}_c \) represents the currents flowing to the capacitor of each \( \pi \)-section line.
\end{itemize}
The state-space matrices for the interconnector are given by:
    
    
%
\begin{equation}
\label{eq.node_interconnector_state_space}
    \begin{aligned}
    \boldsymbol{A}
    &= 
    \begin{bmatrix}
        0
    \end{bmatrix}; 
    & \boldsymbol{B}
    &= 
    \begin{bmatrix}
        0_{1 \times \text{size}(\boldsymbol{D},2)}
    \end{bmatrix}; 
    \\
    \boldsymbol{C}
    &= 
    \begin{bmatrix}
        0_{2 \times 1}
    \end{bmatrix}; 
    & \boldsymbol{D}
    &= 
    \begin{bmatrix}
        \pm1 & 0 & \cdots & \pm1 & 0 \\
        0 & \pm1 & \cdots & 0 & \pm1
    \end{bmatrix};
    \end{aligned}
\end{equation}
where the dimension of the $\boldsymbol{D}$ matrix depends on the number of elements connected to this node, including both lines and shunt elements.
%
%
The sign of the elements ($\pm1$) depends on the current direction of each element.
In \ac{stamp}, the adopted current convention considers as positive the current injection into the node-interconnector. 
\end{definition}

The algorithm uses the incidence matrix of the AC system, combining it with the numerical values of the line parameters, including the values of the cable resistance, inductance, and capacitance.
The definition of the incidence matrix is as follows.
\begin{definition}{(\bf{Incidence matrix}} M) 
of a power system is a matrix of size \( N_{b} \times N_{br} \), where:

\begin{itemize}
    \item $N_{b}$ represents the number of buses (nodes),
    \item $N_{br}$ represents the number of branches (edges).
\end{itemize}

Each entry $m_{ij}$ of $M$ is defined as:

\begin{equation}
m_{ij} = 
\begin{cases} 
      +1 & \text{if branch } j \text{ leaves bus } i, \\
      -1 & \text{if branch } j \text{ enters bus } i, \\
      0 & \text{if branch } j \text{ is not connected to bus } i.
   \end{cases}
\end{equation}
\end{definition}
%
%

For the construction of the AC grid state-space model, \ac{stamp} iterates across the incidence matrix and performs the following operations:
\begin{enumerate}
    \item Computation of the total capacitance for each network bus by aggregating the capacitors from each connected $\pi$-section line.
    \item Calculation of the state-space model of the total capacitance connected to each bus.
    \item Preparing the input names for each node-interconnector to accommodate all the AC system elements.
    \item Calculation of the state-space model for each $RL$-line section.
    \item Calculation of the models for all the node-interconnector subsystems.
    \item Connecting all the state-space models to derive the full state-space model of the AC network.
\end{enumerate}

\subsection{DC grid model}

The DC network is modeled following the same methodology as the AC network.
However, instead of using $\pi$-section line models, the DC case employs a representation consisting of three parallel $RL$ branches, which provides a more accurate representation of the DC line dynamics \cite{Cable_DC0,Cable_DC1,Cable_DCf}.
%

%

Apart from DC lines, other important elements in AC/DC systems are \ac{hvdc} converters, also known as \acp{ipc}~\cite{principles}.
These devices act as an interface between the AC and DC systems and are modeled as hybrid AC/DC admittances $Y_{\rm{ac/dc}}$, seen from both AC and DC grids.
These hybrid admittances are formulated as~\cite{hybrid_admittance}:
\begin{equation}
\label{eq:hybrid_admittance}
    \begin{bmatrix}
        i_q \\
        i_d \\
        i_{\rm{dc}}
    \end{bmatrix} = 
    \underbrace{\begin{bmatrix}
        Y_{qq}(s) & Y_{qd}(s) & Y_{q,\rm{dc}}(s) \\
        Y_{dq}(s) & Y_{dd}(s) & Y_{d,\rm{dc}}(s) \\
        Y_{\rm{dc},q}(s) & Y_{\rm{dc},d}(s) & Y_{\rm{dc}}(s) 
    \end{bmatrix}}_{Y_{\rm{ac / dc}}(s)}
    \begin{bmatrix}
        v_q \\
        v_d \\
        v_{\rm{dc}}
    \end{bmatrix}.    
\end{equation}

$Y_{\rm{ac/dc}}$ can be derived from the \ac{ipc} state-space model, according to the process described in Section~\ref{sec.impmodel} or from impedance estimation methods.
The expected outputs of the \ac{ipc}~(inputs to the AC and DC systems) are the \ac{ipc} currents ($i_{q}$, $i_{d}$, and $i_{\rm{dc}}$), and the expected inputs for the \ac{ipc} element are the voltages from the AC and DC systems ($v_{q}$, $v{_d}$, and $v_{\rm{dc}}$).
\ac{stamp} provides complete models of \acp{ipc} in state-space format with different control roles~\cite{arevalo2024scheduling}.
%


\subsection{Individual Shunt Device Models}
\label{sec.shunt_elements}

\ac{stamp} includes a library of pre-defined linear models, which includes the following components:
\begin{itemize}
    \item \ac{sg}s with different exciters, governors, and mechanical system representations.
    \item \acp{vsc} for AC grid and hybrid AC-DC grid applications with different control algorithms, including \ac{gfm} and \ac{gfl} control configurations.
    \item Ideal grid equivalents, modeled as Thévenin circuits. 
    \item Loads, passive filters, and other shunt elements, modeled as equivalent impedances.
\end{itemize}
%
%
All these elements follow the admittance convention, i.e., they expect the bus voltage at the connection point as the subsystem input, with its current injection to the bus being the output.
Unlike other tools, where the state-space model is computed from symbolic equations, \ac{stamp}
includes the symbolic expressions of the state-space matrices for these elements as integrated functions, allowing their direct numerical evaluation given the system and linearization point parameters, thus avoiding symbolic calculations.

The state-space representation of each shunt device is obtained as the combination of state-space subsystems that represent electric circuits, control blocks, or other algebraic equations, such as the reference frame transformations.
%
%
The names of the input and output variables need to be unique for each subsystem, in order to enable their interconnection at the following stage.
%

\subsection{User-Defined Elements}
\label{sec.user_defined}
Despite \ac{stamp} including an extensive library of pre-defined models, not all possible power system devices and control structures are modeled.
For this reason, a feature to seamlessly incorporate user-defined elements is also considered.
Moreover, sometimes the differential equations of the components are not available due to confidentiality reasons.
%
%
In this case, the state-space models can be provided in numerical form in order to conceal confidential information, or directly estimated from other types of black-box models~\cite{smith2024black,cifuentes2021black}.
%
%
%

%
The incorporation of the user-defined models is implemented in \ac{stamp}.
Upon detecting one instance of such a model, \ac{stamp} reads the
numerical state-space matrices provided by the user.
Assuming an admittance formulation, the tool automatically assigns to the user-defined model states, inputs and outputs names that are compatible with the rest of the system variables, allowing its interconnection with the other state-space subsystems.
These state variables do not have a direct physical interpretation, but can be used to quantify the participation of the user-defined element on the system overall dynamics, e.g., through \ac{pf} analysis.


\section{Complementary Features of the Toolbox}
\label{sec.complementary_features}
Aside from the \ac{ssa} and the state-space model implementation, already presented in Sections~\ref{sec.ssmethods} and~\ref{sec.implementation}, respectively, \ac{stamp} also offers auxiliary features.
The following subsections describe these features that include input data processing, power flow calculation, linearization point calculation, and non-linear model construction and initialization.

\subsection{Input Data Processing}
\label{sec.input_data_processing}
%
%
The input data for \ac{stamp} are defined in a spreadsheet (Microsoft Excel format) and are distinguished between static and dynamic data.
Static data signify the necessary data for the power flow calculation, given a certain operation point.
They include the per-unit bases of the overall system and the individual devices , the network topology and line parameters, as well as the power flow parameters for all the generators and loads of the system.
The power flow parameters for each connected element are comprised of their respective operation point (active and reactive power injection and/or voltage magnitude and angle), as well as the bus type to which they are connected (slack, PV, or PQ).
Finally in this stage, the controller type of each generator is defined.
This includes the exciter and governor type for \acp{sg} and the control mode~(e.g., \ac{gfm}, \ac{gfl}, STATCOM) for \acp{ibr}.
%
%


The dynamic data define the rest of the data that are necessary for the dynamic analysis.
%
These include all the \ac{ibr} and \ac{sg} control parameters as well as various hardware element ratings.


\subsection{Power Flow Calculation}
\label{sec.power_flow}
%
%
The operations of \ac{stamp} start by solving a power flow problem, whose solution is subsequently used in two ways, as illustrated in Fig.~\ref{fig.overview}.
Firstly, it is used for the calculation of the equilibrium point around which the system is linearised.
Secondly, it is used to initialize the states of the non-linear system, thus avoiding computationally inefficient \acp{tds} until the model reaches a steady state~\cite{alican2023initialization}.
%




If the desired power flow solution is already available,
it can be provided directly as input data.
%
Alternatively, \ac{stamp} offers several integrated options for the power flow calculation.
Aside from individually developed routines, an integrated interface with established, open-source software that is tailored for power flow calculations is provided.
Specifically, the user can find designated data parsers and interfaces with MATPOWER and MATACDC for traditional, and AC/DC hybrid power flow calculations, respectively~\cite{matpower,matacdc}.
%
%
%
%

%
\subsection{Linearization point}
\label{sec.linearization}
%


In order to calculate the linearization point of a dynamic system, the equilibrium values of several variables are necessary, aside from the ones that are already provided from the power flow solution.
To calculate these values, \ac{stamp} includes a library of pre-defined functions for each modeled element.
These functions take as input the power flow solution for the interconnection bus of each device, along with the device parameters.
Then, the rest of the required variable values are analytically calculated by solving the linear equations~\eqref{eq:linear_ode} for each device under equilibrium conditions~(i.e., $d \Delta \boldsymbol{x}/dt=0$).
\subsection{Non-linear \ac{tds}}
\label{sec.nonlinear_tds}

For the validation of linear models, \ac{stamp} also includes the feature of automated construction of non-linear, \ac{emt} models in Matlab/Simulink.
%
%
In the following, the process of generating the non-linear models from the data files is explained.
This process consists of the following intermediate steps:
\begin{itemize}
    \item Graph construction: First, a network graph is generated from the data files by using the incidence matrix and the interconnected device list.
    This graph serves as a graphic representation of the interconnections between the system elements. 
    \item Coordinate extraction: Coordinates representing various points within the system are extracted from the constructed graph.
    These coordinates are not meant to represent the actual geographical locations of the installed devices, rather to illustrate the electrical interconnection of the different devices present in the modeled network.
    \item Library of pre-modeled systems: To streamline the modeling process, all included devices are pre-modeled and included in a dedicated Simulink library.
    %
    %
    Each of these models is based on a Simulink mask environment, enabling the code to repeatedly call them from the library when required.
    %
    %
    %
    %
    %
    The model also includes different pre-defined scopes and measurement blocks in order to monitor the operation of the non-linear model during its \ac{tds}.
    \item 
    Non-linear model initialization: \ac{stamp} includes a feature of automatically initializing the non-linear \ac{emt} models.
    %
    %
    %
    The power flow solution is the input for the initialization algorithm with all the initialization variables being computed using predefined Matlab-based functions for each element included in the library~\cite{ding2024holistic}.
    %
    %
    
\end{itemize}


%




\section{Case Studies}
\label{sec.case_studies}

\subsection{WSCC System}
\label{sec.transmission}
\begin{figure}[!t]
\centering
\includegraphics[width=0.85\columnwidth]{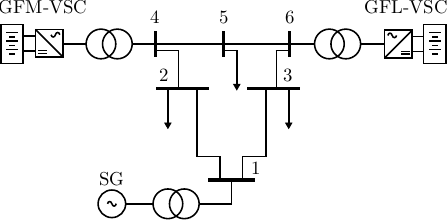}
\vspace{-0.2cm}
\caption{Single-line diagram of the modified WSCC system.}
\label{fig.wscc}
\end{figure} 
\begin{figure}[!h]
\centering
\includegraphics[width=0.75\columnwidth]{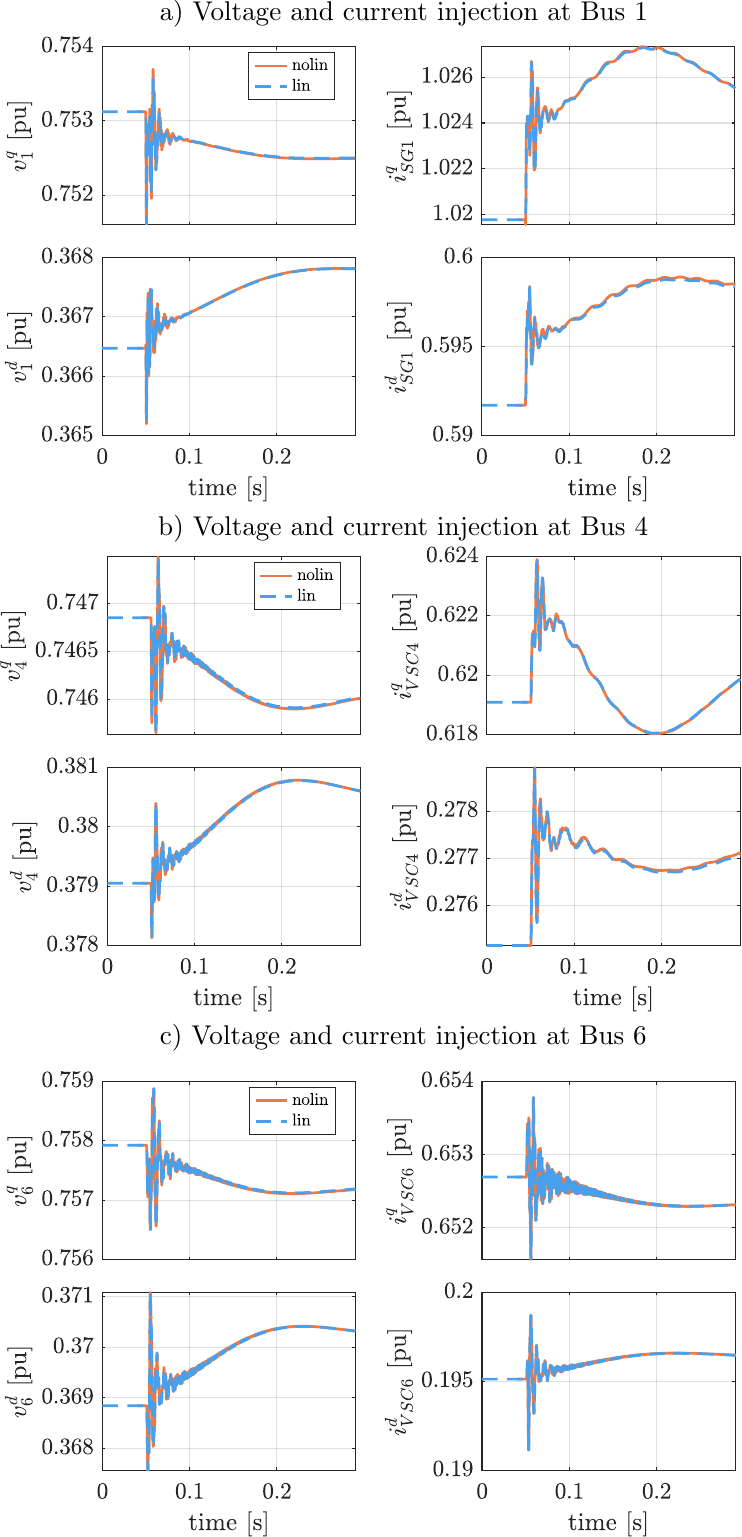}
\vspace{-0.2cm}
\caption{Time domain validation of the WSCC linear model.
}
\label{fig.wscc_tds_validation}
\end{figure} 

In order to illustrate the comprehensive stability analysis capabilities of \ac{stamp}, a case study based on the well-known WSCC system with increased penetration of power converters was performed.
Compared with the original system, two of the standard \acp{sg} were substituted by two \acp{vsc}, one operating in droop-based, \ac{gfm} mode, and one operating in \ac{gfl} mode.
Their control structures and modeling equations are provided in~\cite{pogaku2007modeling,collados}, respectively.
The remaining \ac{sg} was modeled with a standard 6th-order machine model and was equipped with an AC4A exciter and an IEEEG1 turbine governor~\cite{kundur,milano2010power}.
The above generation mix is representative of modern power systems and introduces various dynamic phenomena of different timescales that can be analyzed using \ac{stamp}.
Fig.~\ref{fig.wscc} shows the single-line diagram of the system under study.

The first step in the stability analysis process is the validation of the linear model.
To this end, \ac{tds} is performed for both the linear and non-linear models, both of which are automatically generated by \ac{stamp}, and their dynamics are observed following a perturbation.
Fig.~\ref{fig.wscc_tds_validation} shows the components of the electrical variables~(bus voltage and current injection), expressed in the global rotating reference frame for the generation buses of the system and for a 1\% step increase of the load connected at Bus~2. 
It can be seen that the linear model captures accurately the dynamics of the non-linear model across both the fast and slower timescales in which they evolve. 
\begin{table}[t]
\centering
\caption{WSCC System Eigenvalues with minimum real part and corresponding \ac{pf} analysis}
\begin{tabular}{l|l|l|l|l}
\hline
Eigenvalue & State & \ac{pf} value & State \ac{pf} & \ac{pf} value \\
\hline
0.1977 & $x_{ \rm TURB, 4}$ & 1 & $x_{ \rm TURB, 3}$ & 0.2 \\
\hline
0.8 & $x_{q,v, \rm GFM }$ & 1 & $x_{d,v, \rm GFM}$ & 0.63\\
\end{tabular}
\vspace{+0.2cm}
\label{tab.eig_pfs}
\end{table}

After the linear model validation, the system eigenvalues along with the corresponding \acp{pf} are calculated.
Table~\ref{tab.eig_pfs} shows the two eigenvalues closest to the imaginary axis, along with the two states with the highest \acp{pf} for each eigenvalue, respectively.
It can be seen that for the first eigenvalue, the internal states $x_{ \rm TURB, 4}$ and $x_{ \rm TURB, 3}$ of the \ac{sg} turbine present the highest \acp{pf}, while for the second eigenvalue, the main participating states are the internal states of the \ac{gfm} converter voltage controller, $x_{q,v, \rm GFM }$ and $x_{d,v, \rm GFM }$.
The \ac{pf} analysis indicates which system components are more likely to threaten the system stability and indirectly leads to proper parameter selection for the eigenvalue sensitivity analysis.
\begin{figure}[!t]
\centering
\includegraphics[width=1\columnwidth]{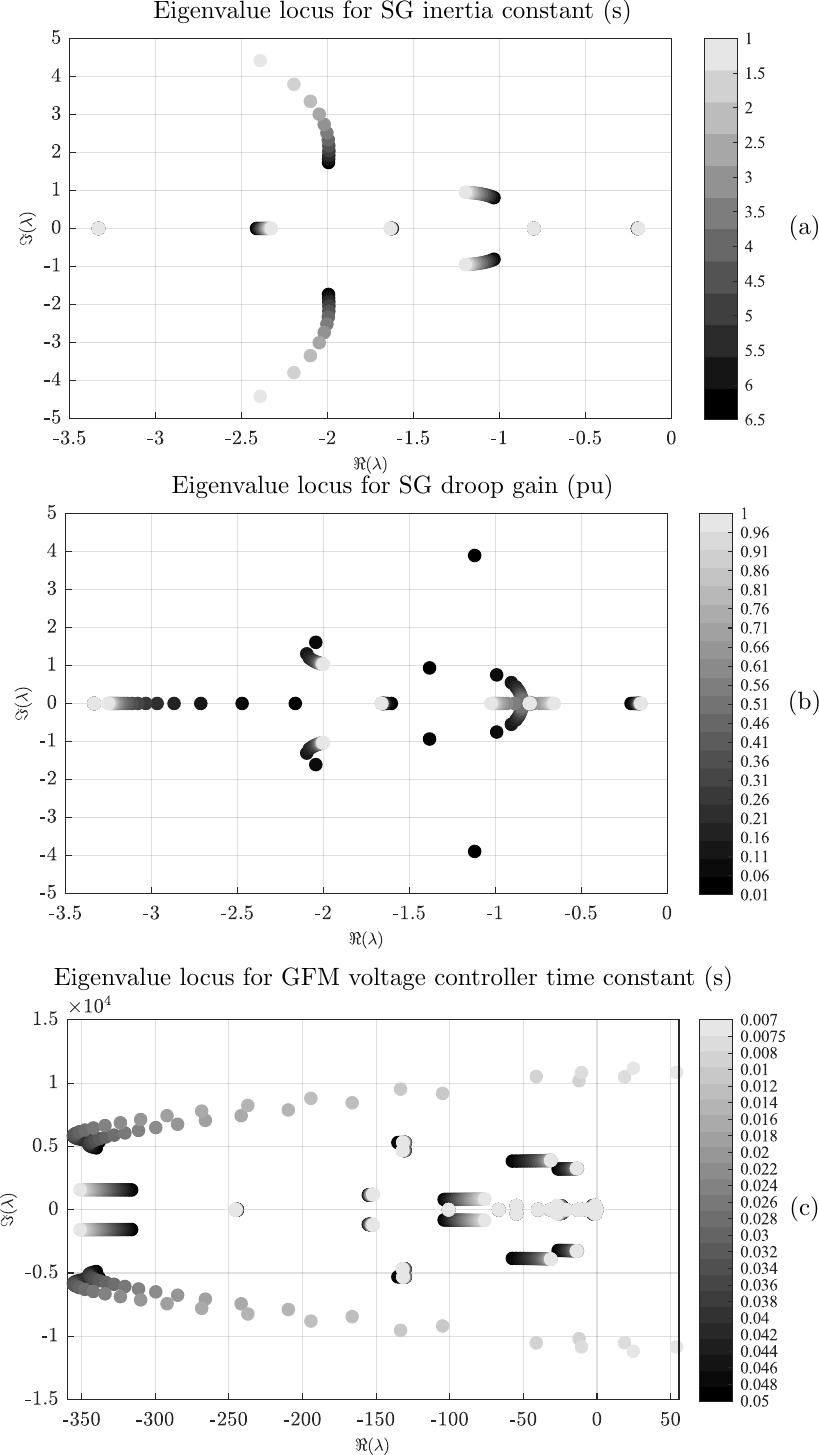}
\vspace{-0.2cm}
\caption{Eigenvalue trajectories for different parameter variations in the WSCC case study. Variation of the (a) \ac{sg} inertia constant, (b) the \ac{sg} droop gain and (c) the \ac{gfm} converter voltage controller time constant.}
\label{fig.wscc_eigenvalue_loci}
\end{figure} 
\begin{figure}[!t]
\centering
\includegraphics[width=1\columnwidth]{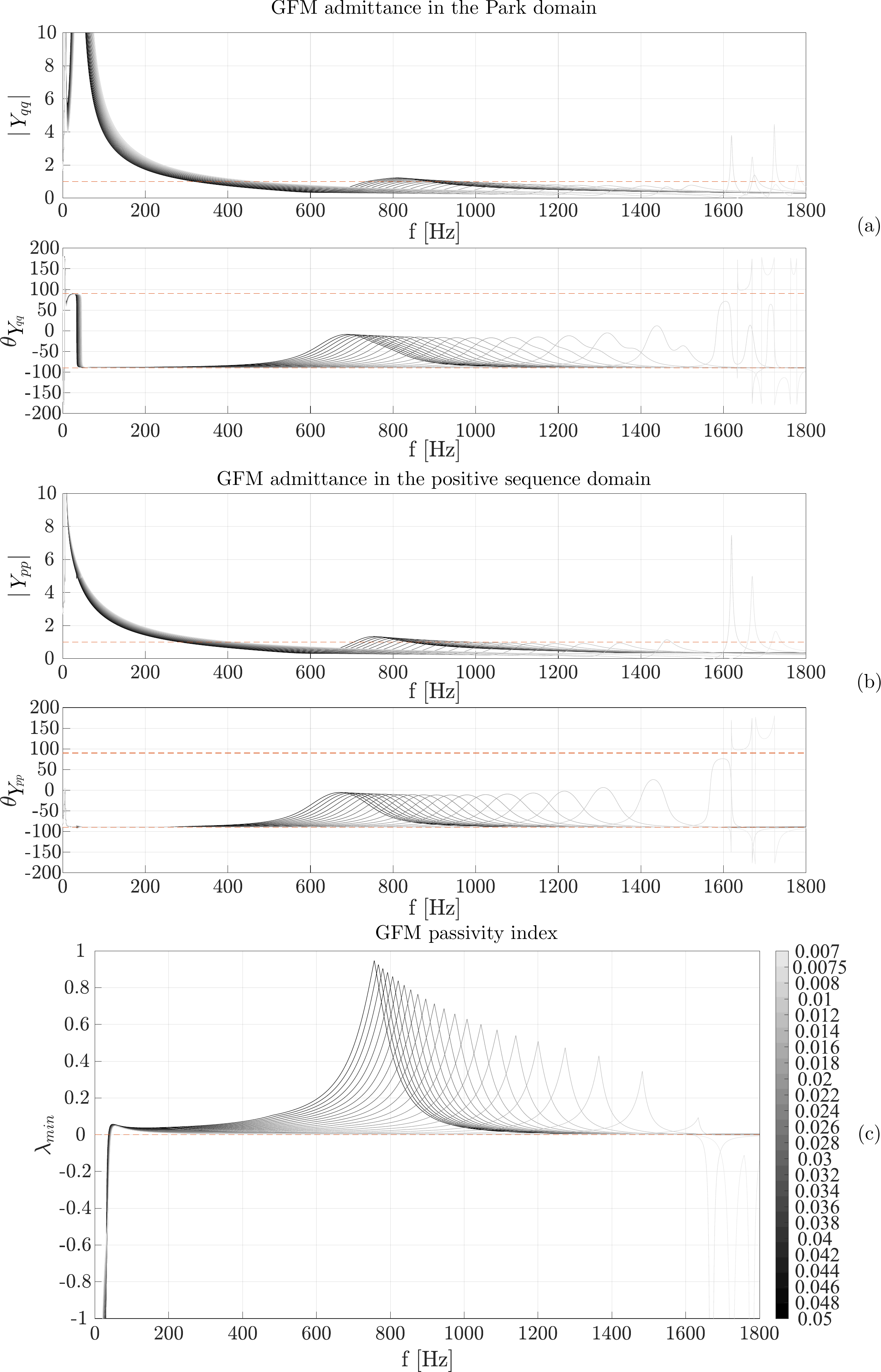}
\vspace{-0.2cm}
\caption{Frequency domain analysis for the \ac{gfm} converter in the WSCC case study and for a variation of the time constant of its voltage controller.
}
\label{fig.wscc_frequency_domain}
\end{figure} 

Fig.~\ref{fig.wscc_eigenvalue_loci} shows the eigenvalue loci for the variation of three system parameters.
Fig.~\ref{fig.wscc_eigenvalue_loci}(a) shows the eigenvalues for a variation of the \ac{sg} inertia constant from the nominal 6.5~s, down to 1~s, while Fig.~\ref{fig.wscc_eigenvalue_loci}(b) shows the eigenvalues for a variation of the \ac{sg} droop constant from 0.01 up to 1.
It can be seen that for both cases, and despite the large parameter space that the variation covers, the eigenvalues maintain a negative real part, verifying that the system remains stable for all the selected parameter values.
Fig.~\ref{fig.wscc_eigenvalue_loci}(c) shows the system eigenvalue trajectories for the variation of the time constant of the \ac{gfm} converter voltage controller from the original value of 50~ms, down to 7~ms.
It can be seen that making the controller faster leads to two pairs of eigenvalues crossing the imaginary axis, rendering the system unstable.
The corresponding frequencies of the two unstable modes are 1726~Hz and 1780~Hz, respectively.

\ac{stamp} offers the possibility of performing parametric stability analysis in the frequency domain.
Fig.~\ref{fig.wscc_frequency_domain} shows the various implemented options to depict the converter operation in the frequency domain, namely the converter admittance expressed in Park and positive sequence coordinates~(in Fig.~\ref{fig.wscc_frequency_domain}(a) and Fig.~\ref{fig.wscc_frequency_domain}(b), respectively), as well as the converter passivity index~(in Fig.~\ref{fig.wscc_frequency_domain}(c)).
The results of the frequency domain analysis are consistent with the eigenvalue analysis, demonstrating that the admittance angle of the converter is outside of the (-90$\degree$,90$\degree$) range in both coordinate systems and for the lowest value of the controller time constant.
Additionally, the frequency region where this occurs coincides with the frequency of the unstable modes of Fig.~\ref{fig.wscc_eigenvalue_loci}(c).
Finally, the passivity plot of Fig.~\ref{fig.wscc_frequency_domain}(c) shows that the converter is non-passive in the same region, implying adverse interactions with the rest of the network in this frequency range and potential instability.
The above frequency domain analysis features allow the seamless incorporation of black-box models to the stability analysis, provided that their respective frequency response is given to the software as an input, as explained in Section~\ref{sec.user_defined}.

\subsection{INELFE System}
\label{sec.hvdc}
\begin{figure}[!t]
    \centering
    \includegraphics[width=1\columnwidth]{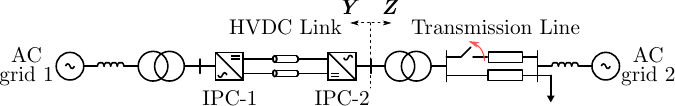}
    \caption{Scheme to represent the instability reported in the INELFE's link in $16^{th}$ June 2015}
    \label{fig:inelfe_scheme}
\end{figure}
\begin{figure}[!t]
\centering
\includegraphics[width=1\columnwidth]{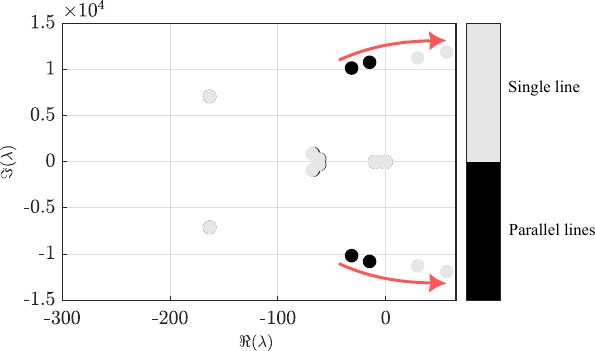}
\vspace{-0.2cm}
\caption{Eigenvalue stability analysis of the INELFE system.
}
\label{fig.inelfe_eig}
\end{figure} 
\begin{figure}[!t]
\centering
\includegraphics[width=1\columnwidth]{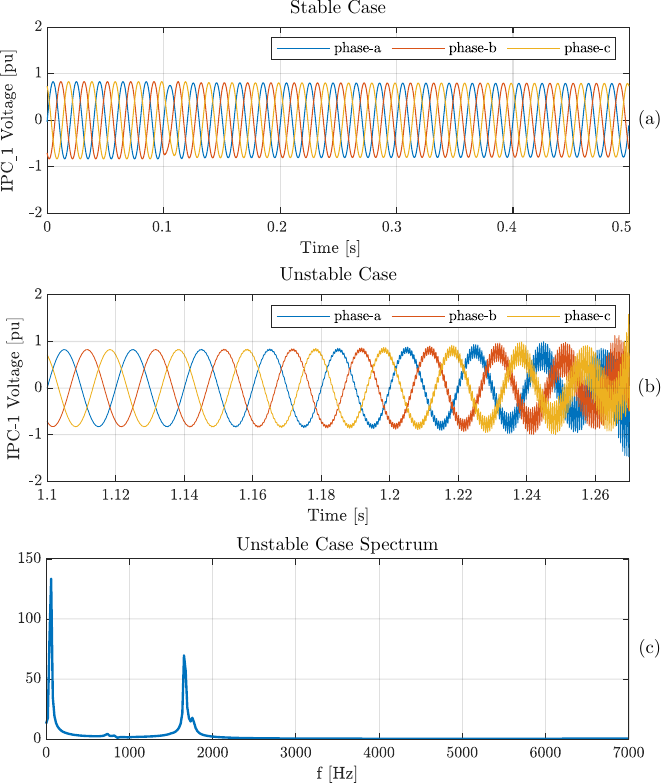}
\vspace{-0.2cm}
\caption{Validation of the INELFE case stability analysis results.
}
\label{fig.inelfe_time}
\end{figure} 

\begin{figure}[t]
    \centering
    \includegraphics[width=1\columnwidth]{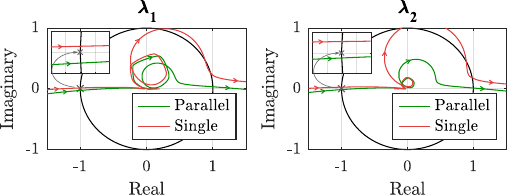}
    \caption{Nyquist general criterion of the INELFE system.
    }
    \label{fig:inelfe_nyquist}
\end{figure}

This case study aims to showcase the capability of \ac{stamp} to perform analysis on AC/DC hybrid power systems
%
and is based on the real-world HVDC interconnection between Spain and France, termed the INELFE link~\cite{francos2012inelfe}.
During the $16^{\text{th}}$ of June 2015, the link suffered a tripping caused from a harmonic instability event, which this case study aims to reproduce~\cite{saad2016performance,saad2017resonances}.

Fig.~\ref{fig:inelfe_scheme} shows the schematic of the system under study.
It is comprised of an HVDC link, two \acp{ipc}, rated at 1000~MW, 400~kV each, a double transmission line, an inductive load, and two Thevenin circuits, representing the two interconnected AC networks. 
\ac{ipc}-1 regulates the DC voltage at the HVDC link while \ac{ipc}-2 controls the active and reactive power interchanged with the AC grid.
%
%
%
%
%
%
The two transmission lines are modeled as $\pi$-section circuits
%
%
and the two Thevenin circuits have the same impedance parameters, namely $R_{th}=0.001$~pu and $X_{th}=0.01$~pu.
The current loop for both \acp{ipc} is tuned with the internal model control method with a time constant of 1~ms~\cite{harnefors1998model,lacerda2022phasor}.
Both \acp{ipc} include a control delay of 50~$\mu$s and a zero-order hold of 10~$\mu$s, modeled as in~\cite{arevalo2024converter}.  


After both the AC and DC power flows are calculated, the system eigenvalues are evaluated using \ac{stamp}.
Fig.~\ref{fig.inelfe_eig} shows the system eigenvalues for two operation scenarios, namely for the case where both transmission lines are connected and in operation~(parallel lines) as well as for the case where one of the lines is disconnected~(single line).
%
%
%
It can be seen that for the parallel lines scenario, the system exhibits a stable operation with all eigenvalues lying in the left half-plane.
For the single line scenario, two unstable modes appear with their eigenvalues being: $\lambda_{1,2} = 56 \pm j11892$ and $\lambda_{3,4} = 29 \pm j11268$, corresponding to frequencies of around 1.9~kHz and 1.8~kHz, respectively.

Fig.~\ref{fig.inelfe_time} shows the validation of the small-signal stability analysis results.
Fig.~\ref{fig.inelfe_time}(a) shows the three-phase voltage at the interconnection point of \ac{ipc}-2 during a 5\% step decrease on the system load at 0.1~seconds.
It can be seen that the \ac{ipc} tracks the new reference resulting from the new operating conditions, maintaining stability.
%
Fig.~\ref{fig.inelfe_time}(b) shows the same variables when the transmission line is disconnected at 0.1~s.
It can be seen that an unstable oscillation appears, leading to a voltage collapse.
Fig.~\ref{fig.inelfe_time}(c) shows the spectrum of the voltage waveform, calculated via a fast Fourier transform algorithm.
The spectral analysis shows a high harmonic content around the frequency of 1.8~kHz, confirming the predicted instability frequency from the small-signal analysis.

The instability phenomenon can also be inspected by the viewpoint of impedance ratio analysis.
To perform this analysis, the system is divided at the interconnection point of \ac{ipc}-2. 
\acp{ipc} 1 and 2, the DC line and the AC grid 1 were represented by an admittance $\boldsymbol{Y}$ in the Laplace domain, while the interconnecting transformer, the double AC transmission line and and the AC grid 2 were represented by an impedance $\boldsymbol{Z}$, as indicated in Fig.~\ref{fig:inelfe_scheme}.
Fig.~\ref{fig:inelfe_nyquist} shows the Nyquist diagram for the two eigenvalues of the minor loop $\boldsymbol{YZ}$ that results from the interconnection of the respective admittance and impedance.
It can be seen that the stability assessment from the impedance criterion is consistent with the one resulting from the modal analysis, with the graph for the single line case encircling the critical point (-1,0), while the one for the parallel line case avoiding it, indicating unstable operation for the former and stable operation for the latter.
\subsection{Case Study Computation Time}
\label{sec.big_system}
\begin{table}[t]
\centering
\caption{Case Study Computation Time Comparison}
\begin{tabular}{|l|l|l|l|l|}
\hline
Case Study & WSCC & INELFE & New England & Midwest
\\
\hline
Number of buses & 6 & 6 & 68 & 113
\\
\hline
Number of states & 88 & 101 & 612 & 1453
\\
\hline
\makecell[l]{Power flow \\ calculation} &
0.081~s
&
0.167~s 
&
0.216~s 
&
0.334~s

\\
\hline
\makecell[l]{Subsystems linear \\ model calculation} 
&
0.441~s
&
0.277~s
 & 
2.044~s
 &
6.242~s
\\
\hline
\makecell[l]{Linear model \\ interconnection} 
&
0.009~s
&
0.008~s
&
0.36~s
&
 2.528~s

\\
\hline
\makecell[l]{Eigenvalue \\ calculation} 
&
0.1~s
&
0.109~s
&
0.616~s 
&
3.895~s

\\
\hline
Total execution &
1.556~s
&
1.505~s 
&
4.774~s
&
16.235~s
\\
\hline
\end{tabular}
\label{tab.computation_time_comparison}
\end{table}
In this Section, the computation time that \ac{stamp} requires its operations is presented.
These operations include the pre-calculation of the power flow for the linearization point evaluation, the calculation of the state space matrices of each individual linear subsystem, their interconnection and the final calculation of the eigenvalues from the complete system state matrix.
The selected systems include the ones of Section~\ref{sec.transmission} and~\ref{sec.hvdc}, as well as of two systems of larger size in order to demonstrate the scalability of the toolbox.
The two larger models are a variation of the well-known New England benchmark, where 35\% of the \acp{sg} are substituted with \acp{ibr}, as well as a reduced version of the Midwest benchmark system.
%
Both the static and dynamic data that were used for the calculations can be found in~\cite{STAMP_Github}. 
All calculations were performed in a desktop computer, equipped with an AMD Ryzen Threadripper 2950X 16-core processor, operating at 3.5 GHz and a RAM of 32 GB.

Table~\ref{tab.computation_time_comparison} shows the the size of each studied system, both in terms of bus and state number, as well as the required computation time for each intermediate step.
It can be seen that the computation times remain at feasible levels, even with a large relative increase of the system size.

\section{Conclusion}
\label{sec.conclusion}
This paper presents~\ac{stamp}, a publicly available, Matlab-based toolbox for comprehensive, \ac{emt}-focused, stability analysis of hybrid AC/DC, converter-based power systems.
The toolbox automatically generates the linear and non-linear models of a given power system
from the specified input data.
By comparing the response in the time domain between the two, the accuracy of the linear model is established.
Subsequently, the linear model can be used for stability-oriented \ac{ssa} with included options for modal, impedance and passivity analyses.
Other complementary features of the toolbox include integrated interfaces with software for AC/DC power flow calculation, linearization point calculation, initialization and \ac{tds} of non-linear \ac{emt} models.

The capabilities of the toolbox are demonstrated and supported with a variety of power system examples of different structure and size.
Future work will focus on expanding the model library included in \ac{stamp} with additional power system devices and controllers, as well as on expanding its capabilities to include options for optimal power flow and fault analysis.
\section*{Acknowledgements}

This work was partially funded by the REFORMING Project (PID2021-127788OA-I00) supported by Fondo Europeo de Desarrollo Regional/Ministerio de Ciencia e Innovación-Agencia Estatal de Investigación. The work of M. Cheah-Mañe and E. Prieto-Araujo is partially funded by the Serra Hunter Programme. O. Gomis-Bellmunt is an ICREA Academic Researcher.

\bibliography{refs_arxiv}
\bibliographystyle{IEEEtran}
\end{document}